\documentstyle[pictex]{article}
\input{psfig}
\begin{document}

\title{
\begin{flushright} {\large INLO-PUB  -- 9  /94} \end{flushright}
\begin{flushright} {\large UWITP  -- 2  /94} \end{flushright}
  \begin{picture}(150,70) \end{picture}\\
  Analytical and numerical methods for massive two-loop self-energy diagrams
  \thanks{Supported by the Stichting FOM and EU contract CHRX-CT-92-0004.}
  \author{S. Bauberger$^{\mathrm{a}}$,
         F. A. Berends$^{\mathrm{b}}$,
         M. B\"ohm$^{\mathrm{a}}$,
         M. Buza$^{\mathrm{b}}$
         \\ \\
         $^{\mathrm{a}}$
         Institut f\"ur Theoretische Physik, Universit\"at W\"urzburg,\\
         Am Hubland, D-97074 W\"urzburg, Federal Republic of Germany \\ \\
         $^{\mathrm{b}}$
         Instituut--Lorentz, University of Leiden,\\
         P.O.B. 9506, 2300 RA Leiden, The Netherlands \\}}
\date{July 1994}
\maketitle
\begin{abstract}
Motivated by the precision results in the electroweak
theory studies of two-loop
Feynman diagrams are performed.
Specifically this paper gives a contribution to
the knowledge of massive two-loop self-energy diagrams in
arbitrary and especially four dimensions.
This is done in three respects: firstly results in terms of
generalized, multivariable
hypergeometric functions are presented giving explicit
series for small and large momenta.
Secondly the imaginary parts of these integrals are expressed as complete
elliptic integrals.
Finally one-dimensional integral representations with elementary functions are
derived.
They are very well suited for the numerical evaluations.
\end{abstract}
\newpage
\section{Introduction}
The beautiful results of the LEP1 experiments have shown that the electroweak
theory has a predictive power like that of QED several decades ago.
It is to be expected
 that eventually the electroweak theory will provide high precision predictions
for many experiments in the present and near future.
One may in particular think
in this respect of the measurement of the Z mass and width, a better
determination of the W mass and a first indication of the value of the
top mass.
This will in the future require two-loop calculations in the electroweak
theory.
One then has to face several problems: amongst others,
a large number of diagrams
and many different non-negligible masses.
It is known that analytical results for
 arbitrary non-vanishing masses are in general not obtainable for two-loop
diagrams in the form of generalized polylogarithms.
The present paper contributes to
 the knowledge about this problem by introducing and applying new analytical
and
 numerical approaches. A number of new results will be presented.

We focus on the simplest class of two-loop diagrams, i.e.
the set of scalar two-loop self-energies.
They are essential for physical predictions and moreover they
show the typical problems one encounters in the evaluation of two-loop
diagrams.
There exist four non-trivial self-energy diagrams
\begin{figure}[htb]
\begin{picture}(450,80)
\put(-50,-490){\includegraphics{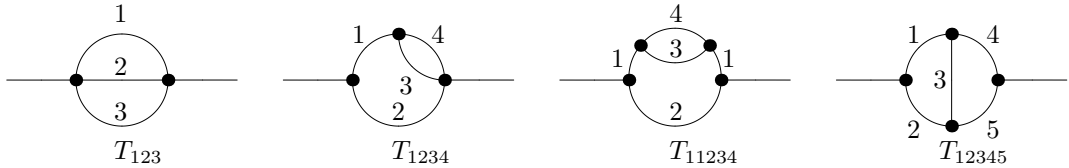}}
\put(60,-12){$T_{123}$}
\put(165,-12){$T_{1234}$}
\put(270,-12){$T_{11234}$}
\put(372,-12){$T_{12345}$}
\put(60,3){$3$}
\put(60,20){$2$}
\put(60,40){$1$}
\put(165,3){$2$}
\put(150,32){$1$}
\put(180,32){$4$}
\put(168,13){$3$}
\put(270,40){$4$}
\put(270,27){$3$}
\put(270,3){$2$}
\put(248,23){1}
\put(290,23){1}
\put(370,15){$3$}
\put(360,32){$1$}
\put(390,32){$4$}
\put(360,-4){$2$}
\put(390,-4){$5$}
\end{picture}
\vspace{3mm}
\caption{Topologies}
\label{fig:top}
\end{figure}

In order to classify our conventions we give the explicit form of the so-called
master diagram:
\begin{equation}
 T_{12345}(p^2;m_1^2,m_2^2,m_3^2,m_4^2,m_5^2) =
 << \frac{1}{k_1^2-m_1^2}\ldots \frac{1}{k_5^2-m_5^2}>>
\end{equation}
where each bracket denotes a $D$ dimensional integration
over one loop momentum $q$:
\begin{equation}
 <\ldots>=\int\frac{d^{D}q}{i\pi^{2}(2\pi\mu)^{D-4}}(\ldots)\,,
\end{equation}
where $\mu$ is an arbitrary mass. The momenta of the propagators
${k_1,\ldots,k_5}$ are determined by momentum
conservation in terms of the external momentum $p$
 and the integration momenta $q_1$ and $q_2$.
It can be immediately seen that
the diagram $T_{11234}(p^2;m_1^2,m_1'^2,m_2^2,m_3^2,m_4^2)$ can be reduced to a
difference of two $T_{1234}$ diagrams by partial fractioning of the propagators
with the same momentum:
\begin{eqnarray}
T_{11234}(p^2;m_1^2,m_1'^2,m_2^2,m_3^2,m_4^2) &=&
\frac{1}{m_1^2-m_1'^2}\left(T_{1234}(p^2;m_1^2,m_2^2,m_3^2,m_4^2)\right.
  \nonumber \\
 && \left.-T_{1234}(p^2;m_1'^2,m_2^2,m_3^2,m_4^2)\right) \,.
\end{eqnarray}
In the special case $m_1=m_1'$ this has to be taken as a derivative with
respect
 to $m_1^2$. So we are only left with three diagrams to be discussed.
Of these
the master diagram is convergent in four dimensions, the other
two are divergent.
 After evaluating them in $D=4-2 \delta$ dimensions,
one would like to derive an
expansion in $\delta$ which starts in $1/\delta^2$ and $1/\delta$ terms.

In recent years several groups derived small and large $p^2$ expansions
for these diagrams.
 For the general massive case they obtained algorithms for symbolic
evaluation of these expansion coefficients. For the general coefficients
in these expansions, no closed expression was derived.
 More recently multiple series were found
 for which the general coefficient is known and which lead to generalized
 hypergeometric functions in several variables.
One aim of this paper is to extend
 these results, an other to express the imaginary parts by elliptic integrals.
 On the other hand an elegant numerical method has been worked out leading to a
two-dimensional integral representation \cite{Kreimer,Czarnecki}.
The second aim of our paper
 is to improve on this by introducing a one-dimensional integral representation
 containing only elementary functions.
For this a general procedure for diagrams
 containing one-loop self-energy expressions in a subloop is presented.
The actual outline of the paper is as follows: in sect. 2 the
analytic approach leading
 to generalized hypergeometric functions and the corresponding explicit series
representations are given, the elliptic integrals for the diagrams are
calculated in sect. 3. The next sect. describes the numerical approach
by means of a one-dimensional
 representation which is derived by a self-energy insertion into one-loop
 diagrams. Finally sect. 5 presents some numerical comparisons
and draws conclusions.
\section{Analytic approaches and hypergeometric functions}
In this section we present analytic results in an arbitrary number of
dimensions
 $D$ for the two-loop scalar self-energy diagrams.
They will be in the form of generalized hypergeometric functions, that is
multiple series of ratios of $p^2$ and the masses.
We also discuss the expansion of these results around $D=4$.

The simplest case of a two-loop scalar self-energy diagram
$T_{123}(p^{2};m_{1}^
{2},m_{2}^{2},m_{3}^{2})$ was discussed in a previous paper \cite{Buza}.
In fact, the result turned out to be represented by Lauricella functions which
could be derived by $x$-space techniques or Mellin-Barnes
representation for a massive propagator.
It turns out that there is a third method using dispersion relations. Since the
latter method will be used several times in this paper we shall first rederive
the result for $T_{123}(p^{2};m_{1}^{2},m_{2}^{2},m_{3}^{2})$
by means of dispersion relations.
Then we focus our attention on the two-loop scalar self-energy diagram with
four
 propagators $T_{1234}(p^{2};m_{1}^{2},m_{2}^{2},m_{3}^{2},m_{4}^{2})$ where
the
small $p^2$ behavior can again be derived by the dispersion method and the
large $p^2$ behavior by the Mellin-Barnes representation method.

Although we have three methods, the master diagram still poses many problems,
the origin of which we will briefly discuss in the last section.

So we start with the rederivation of
$T_{123}(p^{2};m_{1}^{2},m_{2}^{2},m_{3}^{2
})$ by means of dispersion relations. The imaginary part is given by
\begin{eqnarray}
 && Im\left(T_{123}(p^{2};m_1^2,m_2^2,m_3^2)\right)  =
 \frac{\Delta T_{123}(p^{2};m_{1}^{2},m_{2}^{2},m_{3}^{2})}{2 i} \nonumber
 \label{imag}\\
 &&       = - \pi  (4\pi\mu^2)^{2 \delta}
            \frac{\Gamma^2(1-\delta)}{\Gamma^2(2-2 \delta)}
          \Theta\left(p^2-(m_1+m_2+m_3)^2\right)\nonumber\\
 &&  \!\!\! \int\limits _{(m_2 + m_3)^2 } ^{(\sqrt {p^2}-m_1)^2}
        \!\!\! {\mbox{d} s}
        \frac{\lambda^{\frac{1}{2}-\delta} (s,m_2^2,m_3^2)}{s^{1-\delta}}
        \frac{\lambda^{\frac{1}{2}-\delta} (p^2,s,m_1^2)}{{(p^2)^{1-\delta}}}
              \nonumber \\
 && = \frac{1}{4 \pi}\!\!\! \int\limits _{(m_2 + m_3)^2 }^{(\sqrt {p^2}-m_1)^2}
      \!\!\! {\mbox{d} s}
      \Delta B_0(s;m_2^2,m_3^2) \Delta B_0(p^2;s,m_1^2)
\end{eqnarray}
where $\lambda(a,b,c)=(a-b-c)^2-4 b c$ is the K\"{a}ll\'{e}n function.
The dispersion relation reads
\begin{eqnarray}
 \hskip -1mm  T_{123}(p^{2};m_{1}^{2},m_{2}^{2},m_{3}^{2}) &=&
   \frac{1}{2\pi i}\!\!\!
 \!\!\! \int\limits_{(m_{1}+m_{2}+m_{3})^2}^{\infty}\!\!\!\!\! \mbox{d} z
  \frac{1}{z-p^2}
     \Delta T_{123}(z;m_{1}^{2},m_{2}^{2},m_{3}^{2})\label{disp}
     \nonumber \\
 && \hskip -6mm = \frac{1}{4 \pi^2}\int\limits_{(m_{2}+m_{3})^2}^{\infty}\!\!\!
\mbox{d} s
      \,\,\Delta B_0(s;m_2^2,m_3^2) \!\!\!\!
      \int\limits_{(\sqrt s+m_1)^2}^{\infty}\!\!\!\frac{\mbox{d} z}{z-p^2}\,\,
      \Delta B_0(z;s,m_1^2) \,.
\end{eqnarray}
Using the expansion
\begin{equation}
 \frac{1}{z-p^2}=\frac{1}{z} \sum_{k=0}^{\infty} \left(\frac{p^2}{z}\right)^k
\end{equation}
we perform first the integration over $z$:
\begin{eqnarray}
 A &=& \sum_{k=0}^{\infty} (p^2)^k \int\limits_{u}^{\infty} \mbox{d} z\,\,
       z^{\delta-2-k} (z-u)^{1/2-\delta} (z-v)^{1/2-\delta}
      \nonumber \label{A}\\
   &=& \sum_{k=0}^{\infty} (p^2)^k u^{-\delta-k} B(k+\delta,3/2-\delta)
     {}_2 F_1(\delta-1/2,k+\delta;k+3/2;v/u)\,,
\end{eqnarray}
where $u\,=\,(m_1+\sqrt s)^2$ and $v\,=\,(m_1-\sqrt s)^2$.
One transforms the Gauss hypergeometric function ${}_2 F_1$ using
relations which one can find in
\cite{Prudnikov}
\begin{eqnarray}
 &&{}_2 F_1(\delta-1/2,k+\delta;k+3/2;v/u)\!\! =
      \nonumber \label{Ftransf}\\
 && \left(\frac{2 \sqrt s}{m_1+\sqrt s}\right)^{-2(k+\delta)} \!\!\!\!
     {}_2 F_1 (k+\delta,k+1;2 k+2;(1-m_1^2/s))
     \nonumber\\
 && =\left[
     \frac{\Gamma(1-\delta)}{\Gamma(k+2-\delta)}
     {}_2 F_1(k+\delta,k+1;\delta;m_1^2/s)\right.\nonumber\\
 && \!\!\!\left.+\left(\frac{m_1^2}{s}\right)^{1-\delta}
    \frac{\Gamma(\delta-1)}{\Gamma(k+\delta)}
     {}_2 F_1(k+2-\delta,k+1;2-\delta;m_1^2/s)\right]\nonumber\\
 && \times \left(\frac{2 \sqrt s}{m_1+\sqrt s}\right)^{-2(k+\delta)}
    \frac{\Gamma(2 k+2)}{k!}\,.
\end{eqnarray}
With this result, $T_{123}(p^{2};m_{1}^{2},m_{2}^{2},m_{3}^{2})$ becomes:
\begin{eqnarray}
  &&T_{123}(p^{2};m_{1}^{2},m_{2}^{2},m_{3}^{2}) =
     \label{ints} \nonumber \\
  &&-(4\pi\mu^2)^{\delta}
    \frac{\Gamma(1-\delta) \Gamma(\delta-1)}{\Gamma(2-2 \delta)}
    \int\limits_{(m_{2}+m_{3})^2}^{\infty} \mbox{d} s\, s^{\delta-1}
    \lambda^{\frac{1}{2}-\delta} (s,m_2^2,m_3^2)\nonumber\\
  &&\times
    \left[\frac{m_1^2}{s} \left(\frac{m_1^2}{4\pi\mu^2}\right)^{-\delta}
    F_4(1,2-\delta;2-\delta,2-\delta;p^2/s,m_1^2/s)
    \right. \nonumber\\
  &&\left.-\left(\frac{s}{4\pi\mu^2}\right)^{-\delta}
     F_4(1,\delta;2-\delta,\delta;p^2/s,m_1^2/s)\right]\,.
\end{eqnarray}
Since in (\ref{disp}) the $z$ integration up to some factors represents the
one-loop self-energy, we find as a byproduct
\begin{eqnarray}
\label{oneloop}
 B_0(p^2;m_1^2,m_2^2) &=&
    \Gamma(\delta-1)\left[\frac{m_1^2}{m_2^2}\!\!
    \left(\frac{m_1^2}{4\pi\mu^2}\right)^{-\delta}\!\!\!\!
    F_4(1,2\!-\!\delta;2\!-\!\delta,2\!-\!\delta;p^2/m_2^2,m_1^2/m_2^2)
    \right. \nonumber \\
  &&\left.-\left(\frac{m_2^2}{4\pi\mu^2}\right)^{-\delta}
     F_4(1,\delta;2-\delta,\delta;p^2/m_2^2,m_1^2/m_2^2)\right]\nonumber\\
 && =
  \Gamma(\delta)\left(\frac{-p^2}{\mu^2}\right)^{-\delta} \Big\{
         x_2\frac{(x_1 x_2)^{-\delta}}{1-\delta}
         {_2 F_1} (1,\delta,2-\delta;\frac{x_2}{x_1})
                     \nonumber \\
 && \hskip15mm + (1-x_1) \frac{(1-x_1)^{-\delta}(1-x_2)^{-\delta}}
                                                    {1-\delta}
   {_2 F_1} (1,\delta,2-\delta;\frac{1-x_1}{1-x_2})
    \nonumber \\
 && \hskip15mm +(x_1-x_2)^{1-2\delta}
    {\rm B} (1-\delta,1-\delta) (-1)^{-\delta} \Big\},
\end{eqnarray}
where
\begin{equation}
x_{1,2} = \frac{1}{2 p^2} \left( p^2-m_1^2+m_2^2 \pm
\sqrt{\lambda(p^2,m_1^2,m_2^2)}\right)\,.
\end{equation}

Using the definition of the $F_4$ functions we easily perform the integration
over $s$.
After some manipulations similar to the ones used in (\ref{Ftransf}) we obtain
the result for the London transport diagram which agrees with the one derived
using $x$-space techniques or Mellin-Barnes representation
\begin{eqnarray}
&&\hskip-7mm
T_{123}(p^{2};m_{1}^{2},m_{2}^{2},m_{3}^{2}) =
-m_{3}^{2} \left(\frac{m_{3}^{2}}{4\pi\mu^{2}}\right)^{2(\nu-1)}
\times
\label{small} \nonumber \\
&&\hskip-2mm
\left\{
z_{1}^{\nu}z_{2}^{\nu}\;
\Gamma^2(-\nu)\;
F_{C}^{(3)}(1,1+\nu; 1+\nu,1+\nu,1+\nu;z_{1},z_{2},z_{3})
\right.
\nonumber\\
&&\hskip-2mm
-z_{1}^{\nu}\;
\Gamma^2(-\nu)\;
F_{C}^{(3)}(1,1-\nu; 1+\nu,1-\nu,1+\nu;z_{1},z_{2},z_{3})
\nonumber\\
&&\hskip-2mm
-z_{2}^{\nu}\;
\Gamma^2(-\nu)\;
F_{C}^{(3)}(1,1-\nu; 1-\nu,1+  \nu,1+\nu; z_{1},z_{2},z_{3})
\nonumber\\
&&\hskip-2mm
\left.
-\Gamma(\nu)\Gamma(-\nu)\Gamma(1-2\nu)\;
F_{C}^{(3)}(1-2\nu,1-\nu; 1-\nu,1-\nu,1+\nu; z_{1},z_{2},z_{3})
\right\}
\,,
\end{eqnarray}
where $z_{i}=m_{i}^{2}/m_{3}^{2}\,,\,i=1,2\,,z_{3}=p^{2}/m_{3}^{2}$
 and $\nu=1-\delta$.
The above result is valid for small $p^2$ but it can be analytically continued
to the large $p^2$ region, using known analytic continuation formulae for the
Lauricella function $F_{C}^{(3)}$.

With the dispersion method described above we derive the {\sl small} $p^2$
{\sl result} for $T_{1234}$ in $D$
dimensions.
The discontinuity $\Delta T_{1234}$ is a sum of a two and a three-particle cut,
which we denote by $\Delta T_{1234}^{(2)}$ and $\Delta T_{1234}^{(3)}$,
respectively. The two-particle cut is given by
\begin{equation}
 \Delta T_{1234}^{(2)}(p^2;m_i^2) = \Delta B_0(p^2;m_1^2,m_2^2)
  B_0(m_1^2;m_3^2,m_4^2)
 \, ,
\label{DeltaT12342}
\end{equation}
where according to the Cutkosky rules $m_1^2$ in $B_0(m_1^2;m_3^2,m_4^2)$
is considerd as $m_1^2+i \epsilon$, when $m_1 > m_3+m_4$.
Inserting this result in the dispersion integral gives
\begin{eqnarray}
  T_{1234}^{(2)}(p^2;m_i^2) &=& \frac{1}{2\pi i}
                 \int\limits_{(m_1+m_2)^2}^{\infty} \mbox{d} z\,
\frac{1}{z-p^2}
                 \Delta T_{1234}^{(2)}(z;m_i^2)
  \nonumber\\
  &=&B_0(p^2;m_1^2,m_2^2) B_0(m_1^2;m_3^2,m_4^2)\,.
\end{eqnarray}
The discontinuity $\Delta T_{1234}^{(3)}$ is related to the three particle-cut
given in (\ref{imag}) but with an additional factor,
which according to the Cutkosky rules is the complex conjugate
of the propagator $(s-m_1^2+i \epsilon)^{-1}$ and therefore we
get the dispersion relation
\begin{eqnarray}
  &&T_{1234}^{(3)}(p^2;m_i^2) =
    -(4\pi\mu^2)^{2 \delta}
      \frac{\Gamma^2(1-\delta)}{\Gamma^2(2-2 \delta)}
       \int\limits_{(m_{3}+m_{4})^2}^{\infty}\!\! \mbox{d} s\,\,
       \frac{\lambda^{\frac{1}{2}-\delta} (s,m_3^2,m_4^2)}
       {s^{1-\delta}(s-m_1^2-i\epsilon)}\label{T3disp} \nonumber \\
  &&
\phantom{
T_{1234}^{(3)}(p^2) =
        }
\times \int\limits_{(\sqrt s+m_2)^2}^{\infty}\!\! \mbox{d} z
      \frac{\lambda^{\frac{1}{2}-\delta} (s,z,m_2^2)}{z^{1-\delta} (z-p^2)}\,.
\end{eqnarray}
A further discussion of the analytic properties of
$\Delta T_{1234}^{(2)}$, $\Delta T_{1234}^{(3)}$ and
consequently $T_{1234}^{(2)}$, $T_{1234}^{(3)}$ is postponed to subsection 3.2.
After performing the $z$ integration we obtain:
\begin{eqnarray}
&&T_{1234}^{(3)}(p^2;m_i^2) =
-(4\pi\mu^2)^{\delta}
      \frac{\Gamma(1\!-\!\delta)\Gamma(\delta\!-\!1)}{\Gamma(2\!-\!2 \delta)}
 \!\!\! \int\limits_{(m_{3}+m_{4})^2}^{\infty} \!\!\! \mbox{d} s\,
      \frac{s^{\delta-1}}{s-m_1^2} \lambda^{\frac
      {1}{2}-\delta} (s,m_3^2,m_4^2) \nonumber \label{T3s}\\
&&
\phantom{
T_{1234}^{(3)}(p^2) =
        }
\times
    \left[\frac{m_2^2}{s} \left(\frac{m_2^2}{4\pi\mu^2}\right)^{-\delta}
    F_4(1,2-\delta;2-\delta,2-\delta;p^2/s,m_2^2/s)
    \right. \nonumber\\
&&
\phantom{
T_{1234}^{(3)}(p^2) =
        }
  \left.-\left(\frac{s}{4\pi\mu^2}\right)^{-\delta}
     F_4(1,\delta;2-\delta,\delta;p^2/s,m_2^2/s)\right]\,.
\end{eqnarray}

Expanding $(s-m_1^2)^{-1}$ in $m_1^2/s$ and performing the integration  over
$s$
 one gets
\begin{eqnarray}
 &&T_{1234}^{(3)}(p^2;m_i^2) = \frac{\Gamma(1-\delta) \Gamma(1+\delta)}{\delta}
         \left(\frac{m_4^2}{4\pi\mu^2}\right)^{- 2\delta}
                   \label{T3} \nonumber\\
 &&
\times \sum_{m,n,k,l=0}^{\infty} z_1^k z_2^n (1-z_3)^l z_4^m
\frac{\Gamma(1+m+n)\Gamma(1+\delta+m+n+k+l)}{\Gamma(2-\delta+m)
      m!n!l!}\nonumber\\
 &&
\phantom{
\sum_{m,n,k,l=0}^{\infty}
        }
\left(z_2^{1-\delta}
\frac{\Gamma(2-\delta+m+n)\Gamma(2+m+n+k+l)}
     {\Gamma(2-\delta+n)\Gamma(2(m+n+k)+4+l)}\right.\nonumber\\
 &&
\phantom{
\sum_{m,n,k,l=0}^{\infty}
        }
     \left.-\frac{\Gamma(\delta+m+n)\Gamma(2 \delta+m+n+k+l)}
           {\Gamma(\delta+n)\Gamma(2(m+n+k)+2+2 \delta+l)}
           \right)\,,
\end{eqnarray}
where $z_i\,=\,m_i^2/m_4^2$ with $i=1,2,3$ and $z_4\,=\,p^2/m_4^2$.
Note that the contribution from the three-particle cut is written
in terms of multiple series which are no longer Lauricella functions
but rather belong to a special class of generalized
hypergeometric functions.
To our knowledge they have not been studied in the mathematical literature.
One can however obtain information on the convergence region.
The series in $(1-z_3)$ can be written as ${}_2 F_1$ functions which can
be transformed into ${}_2 F_1$ functions with variable $z_3$.
One then obtains four quartic series in $z_i$. Applying the standard
reasoning (see e.g. \cite{Exton}) we get the following conditions for
convergence
\begin{equation}
m_2 + \sqrt {|p^2|} <  m_1 \,\,\,\,\,\,
\mbox{and}\,\,\,\,\, m_1 + m_3 <  m_4.
\end{equation}

With these results $T_{1234}$ becomes
\begin{equation}
T_{1234}(p^{2};m_i^2) =
        B_0(p^2;m_1^2,m_2^2) B_0(m_1^2;m_3^2,m_4^2)
        +T_{1234}^{(3)}(p^2;m_i^2)\,.
       \label{T1234}\\
\end{equation}

Next the general small $p^2$ result for $T_{1234}$ will be expanded in
$\delta$.

 The following combination of the general massive case with a massless case is
chosen in such a way that the infinite parts cancel  \cite{Berends}
\begin{eqnarray}
\label{finite}
 T_{1234N}(p^2;m_1^2,m_2^2,m_3^2,m_4^2) & = &
 T_{1234}(p^2;m_1^2,m_2^2,m_3^2,m_4^2)   \nonumber \label{TN}\\
 &&- T_{1234}(p^2;m_1^2,m_2^2,0,0)\,.
\end{eqnarray}

An analytic form of this combination is obtained by expanding the multiple
series and their coefficients in $\delta$, where the first and the second
logarithmic derivatives of the $\Gamma$-function occur at integer arguments
\begin{equation}
\psi(n+1)=-\gamma+\sum_{k=1}^{n}\frac{1}{k},\,\,\,\,\,\,\,\,
\psi'(n+1)=\zeta(2)-\sum_{k=1}^{n}\frac{1}{k^{2}}\,,
\end{equation}
with the Euler constant $\gamma$ and $\zeta(2)=\pi^{2}/6$.

The $1/\delta^2$, $1/\delta$ and $\gamma$ terms indeed drop out
from the result
and a finite combination of various multiple series remains and some other
terms
 corresponding to the finite parts of $T_{1234}(p^2;m_1^2,m_2^2,0,0)$ and
$B_0(p^2;m_1^2,m_2^2) B_0(m_1^2;m_3^2,m_4^2)$
 which are given in \cite{Scharf}.
The result is given in the appendix.


The {\sl large} $p^2$ {\sl result} has been derived using the Mellin-Barnes
representation for a massive propagator
\begin{equation}
\frac{1}{(k^{2}-m^{2})^{\alpha}}=
\frac{1}{\Gamma(\alpha)}\,
\frac{1}{2\pi i}
\int\limits_{-i\infty}^{+i\infty}ds\,
\frac{(-m^{2})^{s}}{(k^{2})^{\alpha+s}}
\Gamma(-s)\Gamma(\alpha+s)\,,
\label{Mellin}
\end{equation}
where the integration contour in the s plane must separate
the series of poles of $\Gamma(-s)$ on the right
from the series of poles of $\Gamma(s+\alpha)$ on the left.
In the expression for $T_{1234}$ we
apply (\ref{Mellin}) with $\alpha=1$
to all propagators, thereby relating the general massive case
to the massless one, but with the
propagators raised to arbitrary powers
$1\!+\!s_{i}$.
The required expression is well-known, see e.g.\cite{Smirnov}
\begin{eqnarray}
 &&\hskip-5mm T_{1234}(p^{2};0,0,0,0;1\!+\!s_{1},1\!+\!s_{2},1\!+\!s_{3},
                                     1\! +\!s_4)=
\vphantom{\frac{\Gamma}{\Gamma}}
 \frac{e^{-i\pi \sum_{i=1}^{4} s_i}}{(4\pi\mu^2)^{D-4}}\,
(-p^{2})^{D-4-\sum_{i=1}^{4} s_i} \nonumber \\
 &&\hskip-5mm
\times
\frac{
     \Gamma(2+s_3+s_4-D/2)
     \Gamma(D/2-1-s_{2})\Gamma(D/2-1-s_{3})\Gamma(D/2-1-s_{4})}
     {\Gamma(1+s_{2})\Gamma(1+s_{3})\Gamma(1+s_{3})
     \Gamma(D-2-s_3-s_4)
     } \nonumber\\
 && \times
\frac{\Gamma(4-D+\sum_{i=1}^{4} s_i)\Gamma(D-3-s_1-s_2-s_3)}
{\Gamma(3D/2-4-\sum_{i=1}^{4} s_i)\Gamma(3+s_1+s_3+s_4-D/2)}.
\end{eqnarray}
Closing the integration contours in a way that the convergence is
guaranteed we get
\begin{equation}
\label{large}
    T_{1234}(p^{2};m_{1}^{2},m_{2}^{2},m_{3}^{2},m_{4}^2) =
    \left(\frac{-p^2}{4\pi\mu^{2}}\right)^{2\nu-2}
    \left\{\sum_{i=1}^{6} c_i B_i + B\right\}
\end{equation}
where
\begin{eqnarray}
    B  &=& \left(-x_4\right)^{2\nu-1}\,\,\,
           \frac{\Gamma(1-\nu)}{\Gamma(1+\nu)}
           F_4(1,1-\nu;1+\nu,1-\nu;x_1,x_2) \nonumber \\
 && \left[ \Gamma^2(\nu) \Gamma(1- 2 \nu)
     F_4(1-\nu,1-2 \nu;1-\nu,1-\nu;r,t) \right. \nonumber \\
 && + t^{\nu} \Gamma(\nu) \Gamma(-\nu)
     F_4(1,1- \nu;1-\nu,1+\nu;r,t) \nonumber \\
 && + r^{\nu} \Gamma(\nu) \Gamma(-\nu)
     F_4(1,1- \nu;1+\nu,1-\nu;r,t) \nonumber \\
 && \left.+ r^{\nu} t^{\nu}
    \frac{\Gamma^2(-\nu)\Gamma(1+\nu)}{\Gamma(1-\nu)}
     F_4(1,1+\nu;1+\nu,1+\nu;r,t) \right]
\end{eqnarray}
with $x_i\,=\,m_i^2/p^2$ with $i\,=\,1,2,3,4$, $r\,=\,m_1^2/m_4^2$
and $t\,=\,m_3^2/m_4^2$.
 Between the brackets we recognize up to some factors the result for the vacuum
diagram with three massive lines \cite{Davyd}. Furthermore
the coefficients $c_i$ are
\begin{eqnarray}
c_1 &=& \frac{ \Gamma^2(\nu) \Gamma(\nu-1)\Gamma(1-2\nu)}{\Gamma(3 \nu-1)}
        \nonumber \,, \\
c_2 &=& (-x_2)^{\nu}\frac{\Gamma(\nu)\Gamma(-\nu)\Gamma(\nu-1)\Gamma(2-\nu)}
                      {\Gamma(2\nu-1)}\nonumber \,, \\
c_3 &=& (-x_3)^{\nu}\frac{\Gamma(\nu)\Gamma(-\nu)\Gamma(\nu-1)\Gamma(2-\nu)}
                      {\Gamma(2\nu-1)}\nonumber \,, \\
c_4 &=& (-x_4)^{\nu}\frac{\Gamma^2(\nu)\Gamma(-\nu)\Gamma(1-\nu)}{\Gamma(2\nu)}
        \nonumber \,, \\
c_5 &=& (-x_2)^{\nu} (-x_4)^{\nu}\Gamma^2(-\nu) \nonumber \,,  \\
c_6 &=& (-x_2)^{\nu} (-x_3)^{\nu}\Gamma^2(-\nu) \nonumber \,.
\end{eqnarray}
The other terms are quartic series

\begin{eqnarray}
 B_1 \!\!\!&=&\!\!\! \sum_{j_1=0}^{\infty}\!\ldots\!\!\sum_{j_4=0}^{\infty}
     \frac{(1)_{j_1} (1-\nu)_{j_3+j_4} (1-2\nu)_{j_3+j_4}
           (2-2\nu)_{j_1+j_2+j_3+j_4} (2-3\nu)_{j_1+j_2+j_3+j_4}}
          {(1-\nu)_{j_2} (1-\nu)_{j_3} (1-\nu)_{j_4}
           (2-\nu)_{j_1+j_3+j_4} (2-2\nu)_{j_1+j_3+j_4}} q\,,
 \nonumber \\
 B_2 \!\!\!&=&\!\!\! \sum_{j_1=0}^{\infty}\!\ldots\!\!\sum_{j_4=0}^{\infty}
     \frac{(1)_{j_1} (1-\nu)_{j_3+j_4} (1-2\nu)_{j_3+j_4}
           (2-2\nu)_{j_1+j_2+j_3+j_4} (2-\nu)_{j_1+j_2+j_3+j_4}}
          {(1+\nu)_{j_2} (1-\nu)_{j_3} (1-\nu)_{j_4}
           (2-\nu)_{j_1+j_3+j_4} (2-2\nu)_{j_1+j_3+j_4}} q\,,
 \nonumber \\
 B_3 \!\!\!&=&\!\!\! \sum_{j_1=0}^{\infty}\!\ldots\!\!\sum_{j_4=0}^{\infty}
     \frac{(1)_{j_1} (1)_{j_3+j_4} (1-\nu)_{j_3+j_4}
           (2-2\nu)_{j_1+j_2+j_3+j_4} (2-\nu)_{j_1+j_2+j_3+j_4}}
          {(1-\nu)_{j_2} (1+\nu)_{j_3} (1-\nu)_{j_4}
           (2-\nu)_{j_1+j_3+j_4} (2)_{j_1+j_3+j_4}} q\,,
 \nonumber \\
 B_4 \!\!\!&=&\!\!\! \sum_{j_1=0}^{\infty}\!\ldots\!\!\sum_{j_4=0}^{\infty}
     \frac{(1)_{j_1} (1)_{j_3+j_4} (1-\nu)_{j_3+j_4}
           (2-2\nu)_{j_1+j_2+j_3+j_4} (2-\nu)_{j_1+j_2+j_3+j_4}}
          {(1-\nu)_{j_2} (1-\nu)_{j_3} (1+\nu)_{j_4}
           (2-\nu)_{j_1+j_3+j_4} (2)_{j_1+j_3+j_4}} q \,,
 \nonumber \\
 B_5 \!\!\!&=&\!\!\! \sum_{j_1=0}^{\infty}\!\ldots\!\!\sum_{j_4=0}^{\infty}
     \frac{(1)_{j_1} (1)_{j_3+j_4} (1-\nu)_{j_3+j_4}
           (2)_{j_1+j_2+j_3+j_4} (2-\nu)_{j_1+j_2+j_3+j_4}}
          {(1+\nu)_{j_2} (1-\nu)_{j_3} (1+\nu)_{j_4}
          (2-\nu)_{j_1+j_3+j_4} (2)_{j_1+j_3+j_4}} q\,,
 \nonumber\\
 B_6 \!\!\!&=&\!\!\! \sum_{j_1=0}^{\infty}\!\ldots\!\!\sum_{j_4=0}^{\infty}
     \frac{(1)_{j_1} (1)_{j_3+j_4} (1-\nu)_{j_3+j_4}
           (2)_{j_1+j_2+j_3+j_4} (2-\nu)_{j_1+j_2+j_3+j_4}}
          {(1+\nu)_{j_2} (1+\nu)_{j_3} (1-\nu)_{j_4}
           (2-\nu)_{j_1+j_3+j_4} (2)_{j_1+j_3+j_4}} q\,,
\end{eqnarray}
where
\begin{equation}
 q = \frac{x_1^{j_1} x_2^{j_2} x_3^{j_3} x_4^{j_4}}{j_1! j_2! j_3! j_4!}\,.
\end{equation}

The large $p^2$ result given by (\ref{large}) is valid for
\begin{equation}
 |p^2| > (m_2+m_3+m_4)^2 \,\,\,\, \mbox{and} \,\,\,\,\, m_1 > m_3+m_4\,.
\end{equation}

One may wonder what the relation is between the large $p^{2}$ expansion
of (\ref{large}) and that given in \cite{Smirnov}. In the latter
approach the various terms in the $p^{2}$ expansion are obtained
from the expansion of subgraphs. The subgraphs are obtained by
distributing the momentum $p$ over the propagators in all possible ways.
In the case of the diagram $T_{1234}$ one has the
following subgraphs: the diagram itself, the four diagrams where
one internal line is removed, the two diagrams where two internal lines are
removed and one diagram where three internal lines are removed.
So one expects eight subgraphs, in fact seven because the contribution from the
one where the line corresponding to propagator 1 in our convention is removed
is
 zero.

Following the analysis of \cite{Smirnov} one can easily find the
first term of each of the contributing
series. For the subgraph representing the whole diagram the first term in
the series should be the massless diagram. This series then
corresponds to the first term in (\ref{large}), i.e. $B_1$.
The series which originates from the subgraph where two lines
have been removed, e.g. 2 and 3, starts
with the product of two massive tadpoles. They contribute a
factor $\left(m_{2}^{2}m_{3}^{2}\right)^{\nu}$
which can be identified with the sixth term in (\ref{large}).
The remaining subgraphs
are obtained by removing one internal line, e.g. line 3.
This yields a series starting with a massive
tadpole proportional to $(m_{3}^{2})^{\nu}$.
This is the third term in (\ref{large}).
The last term in (\ref{large}) corresponds to the diagram where three internal
lines are removed whose reduced graph is the vacuum diagram.
Thus the seven series in (\ref{large}) can be related
directly to the seven subgraphs
which are required for the method of \cite{Smirnov}.

The large $p^2$ expansion satisfies a system of four partial
differential equations given below
\begin{equation}
\left({\cal{T}}_1 (1+D_1) x_1^{-1} - {\cal{T}}_2 (1+D_1) \right)f = 0 \,,
\end{equation}
\begin{equation}
\left((d+D_2)(1+ D_2)x_2^{-1} - {\cal{T}}_2 \right)f = 0 \,,
\end{equation}
\begin{equation}
\left( {\cal{T}}_1 (d+ D_3)(1+ D_3)x_3^{-1} - {\cal{T}}_2
        (d+ D_3+ D_4)(e+ D_3+ D_4)\right)f = 0 \,,
\end{equation}
\begin{equation}
\left( {\cal{T}}_1 (d+ D_4)(1+ D_4)x_4^{-1} - {\cal{T}}_2
        (d+ D_3+ D_4)(e+ D_3+ D_4)\right)f = 0 \,,
\end{equation}
where
\begin{equation}
{\cal{T}}_1 = (a+\!\!\sum_{i=1,3,4}\!\! D_i)(c+\!\!\sum_{i=1,3,4}\!\! D_i)\,,
\end{equation}
\begin{equation}
{\cal{T}}_2 = (a+\sum_{i=1}^{4} D_i)(b+\sum_{i=1}^{4} D_i)\,,
\end{equation}
$a=(2-2 \nu)$, $b=(2-3 \nu)$, $c=(2-\nu)$, $d=(1-\nu)$,
$e=(1-2 \nu)$ and $D_i=x_i \partial/ \partial x_i$.

The expressions for small and large $p^2$ are given in an arbitrary
number of dimensions and for arbitrary masses.
One may subsequently derive {\sl special cases} setting masses equal to zero.
Taking the small $p^2$ expansion, $m_3$ and $m_4$ cannot be taken
zero simultaneously.
When enough masses are zero and others are equal it is not so difficult
to recognize known hypergeometric functions and to perform an expansion
in $\delta$.
We give the following examples.

We will use the following abbreviations
\begin{equation}
L_m = \gamma + \ln \left(\frac{m^2}{4\pi\mu^2}\right) ,\,\,\,\,\,\,
L_p = \gamma + \ln \left(\frac{-p^2}{4\pi\mu^2}\right)\,.
\end{equation}

In the case $m_1=m_2=m_3=0$ the three-particle cut contribution (\ref{T3})
reads
\begin{equation}
  T_{1234}^{(3)}(p^2;0,0,0,m^2) =
 \Gamma(1\!-\!\delta)\Gamma(\delta\!-\!1)\Gamma(2 \delta)\,
  {}_3 F_2 (1,\delta,2 \delta;2,2\!-\!\delta;x)
\end{equation}
and the two-particle cut contribution is given by
\begin{equation}
 B_0(p^2;0,0)B_0(0;0,m^2) = -\left(\frac{-p^2}{4\pi\mu^2}\right)^{-\delta}
                             \left(\frac{m^2}{4\pi\mu^2}\right)^{-\delta}
                    \frac{\Gamma^2(1-\delta)\Gamma(\delta)\Gamma(\delta-1)}
                         {\Gamma(2-2 \delta)}\,.
\end{equation}
Expanding the two results from above in $\delta$ we obtain
\begin{eqnarray}
 T_{1234}(p^2;0,0,0,m^2) &=&
   \frac{1}{2 \delta^{2}} + \frac{1}{\delta}
 \left\{ \frac{5}{2} - L_m  - \ln \left(-x\right) \right\}
  \nonumber \\
 &&+ \frac{19}{2} - \frac{3}{2} \zeta(2) + L_m^2
+ \left( - 5 + 2 \ln \left(-x\right) \right) L_m
+ \frac{1}{2} \ln^2 \left( -x\right)
 \nonumber\\
 &&- 3\, \ln \left( -x\right)
 - 2 \frac{x-1}{x} \ln \left( 1 - x\right) - \frac{x+1}{x} \rm{Li_{2}}(x)\,,
\end{eqnarray}
where $x=p^2/m^2$, in agreement with \cite{Scharf}.

In the case $m_1=m_2=0$ and $m_3=m_4=m$ the contribution from the
three-particle
 cut (\ref{T3}) becomes
\begin{eqnarray}
 T_{1234}^{(3)}(p^2) &=&\!\!
 2 \frac{\Gamma(\delta\!-\!1) \Gamma(2 \delta) \Gamma(2\!+\!\delta)}
        {\Gamma(3\!+\!2 \delta)}
   \left(\frac{m^2}{4\pi\mu^2}\right)^{-2 \delta} \!\!\!\!\!\!
  {}_3 F_2(1,\delta,2 \delta;2\!-\!\delta,3/2\! +\!\delta;x/4)
  \nonumber  \\
 &=& -\frac{1}{2 \delta^{2}} + \frac{1}{\delta}
 \left(\frac{1}{2} +  L_m\right) -L_m^2 -L_m -\frac{3}{2} -
     \frac{1}{2} \zeta(2)
      + Y + O(\delta) \,,
\end{eqnarray}
where
\begin{eqnarray}
 Y &=& - \sum_{n=1}^{\infty} \frac{\Gamma^2(n)}{(1+n) \Gamma(2 n+2)} x^n
    \nonumber \\
 &=&  - \frac{x}{6} ({}_3 F_2(1,1,1;2,5/2;x/4)
      -\frac{1}{2}{}_3  F_2(1,1,2;3,5/2;x/4))\,.
\end{eqnarray}
Knowing the analytic expression for the $B_0$ functions which occur in
the two-particle cut contribution and expanding in $\delta$ we obtain
\begin{eqnarray}
 T_{1234}(p^2;0,0,m^2,m^2) &=&
          \frac{1}{2 \delta^{2}} + \frac{1}{\delta}
 \left(\frac{5}{2} - L_p\right) + \frac{19}{2} - \frac{1}{2} \zeta(2)
          \nonumber\\
 && + L_p^2 - 5 L_p + 3 \ln \left(\frac{-p^2}{m^2} \right) -
      \frac{1}{2} \ln^2 \left(\frac{-p^2}{m^2} \right) \nonumber\\
 && + 3 \frac{m^2}{p^2}(r_1-r_2) \ln(r_1)
    + \frac{1}{2}\left(1+\frac{2 m^2}{p^2}\right) \ln^2(r_1)\,,
\end{eqnarray}
where $r_1$ and $r_2$ are the roots of the equation
\begin{equation}
 m^2 r\, +\, \frac{m^2}{r}\, =\, 2 m^2 \,-\, p^2\,\,,
\end{equation}
in agreement with \cite{Berends}.

When $m_2=m_3=0$ and $m_1=m_4=m$ eq. (\ref{T3}) gives
\begin{equation}
 T_{1234}^{(3)}(p^2;m^2,0,0,m^2) =
 -\frac{\Gamma(\delta) \Gamma(2 \delta) \Gamma(1-\delta)}
       {(1-\delta)(1-2 \delta)}\left(\frac{m^2}{4\pi\mu^2}\right)^{-2 \delta}
 \!\!\!
 {}_2 F_1(2 \delta,\delta;2-\delta;x)\,.
\end{equation}
To this the contribution $B_0(p^2;m^2,0)B_0(m^2;0,m^2)$ should be added
\begin{equation}
 T_{1234}^{(2)}(p^2;m^2,0,0,m^2) =
 \frac{\Gamma^2(1+\delta)}{\delta^2 (1-\delta) (1-2 \delta)}
 \left(\frac{m^2}{4\pi\mu^2} \right)^{-2 \delta}
 {}_2 F_1(1,\delta;2-\delta;x)\,.
\end{equation}
Expanding in $\delta$ we get
\begin{eqnarray}
 T_{1234}(p^2;m^2,0,0,m^2) &=&
 \frac{1}{2 \delta^{2}}+ \frac{1}{2 \delta}
 \left\{5 - 2 L_m+ 2 \left(\frac{1-x}{x}\right)
 \ln \left(1 - x\right)\right\}  \nonumber\\
 &&
 + \frac{19}{2} - \frac{1}{2} \zeta(2) + L_m^2 - 5 L_m
 -2 L_m \left(\frac{1-x}{x}\right) \ln \left(1 - x\right)\nonumber\\
 &&
 + 5 \left(\frac{1-x}{x}\right) \ln \left(1 - x\right)
 - \left(\frac{1-x}{x}\right) \ln^2 \left(1 - x\right)\nonumber\\
 &&
 - \frac{1}{x} \rm{Li_2}(x) \,.
\end{eqnarray}

In principle there is one class of cases which has not yet been covered by the
above multiple series.
That is the case where $m_3=m_4=0$. For completeness we give the
small $p^2$ case as a linear combination of an $F_2$ and an $F_4$
 functions \cite{Exton,Appell,Srivastava}. For the derivation we used
the Mellin-Barnes representation method.
\begin{eqnarray}
 T_{1234}(p^2;m_1^2,m_2^2,0,0) &=&
          \left(\frac{m_2^2}{4\pi\mu^2}\right)^{-2\delta}
          \Gamma(1-\delta)\Gamma(\delta-1)\Gamma(2 \delta-1)
         \nonumber \\
 && \left[
         (1 \pm \sqrt y)^{-2(1+\delta)}
         F_2 (1+\delta,1,3/2-\delta;1+\delta,3-2 \delta;z_1,z_2)\right.
         \nonumber \\
 && \left.- x^{1-2 \delta}
    F_4(1,2-\delta;2-\delta,2-\delta;x,y)
    \right]
\end{eqnarray}
where $x\,=\,m_1^2/m_2^2$, $y\,=\,p^2/m_2^2$, $z_1\,=\,x/(1 \pm \sqrt y)^2$
and $z_2\,=\,\pm 4 \sqrt y/(1 \pm \sqrt y)^2$.

As it was mentioned in the introduction the case $T_{11234}$ with
$m_1'=m_1$ can be obtained from $T_{1234}$ by differentiating with
respect to $m_1^2$.
For instance, differentiating eq. (\ref{T1234}) leads to
\begin{eqnarray}
&& T_{11234}(p^2;m_1^2,m_1^2,m_2^2,m_3^2,m_4^2) =
\frac{\partial}{\partial m_1^2} T_{1234}^{(3)}(p^2;m_1^2,m_2^2,m_3^2,m_4^2)
\label{T11234} \nonumber  \\
&& + B_0(p^2;\nu_1=2,\nu_2=1;m_1^2,m_2^2) B_0(m_1^2;m_3^2,m_4^2) \nonumber \\
&& + B_0(p^2;m_1^2,m_2^2) \frac{\partial}{\partial m_1^2}
     B_0(m_1^2;m_3^2,m_4^2) \,.
\end{eqnarray}
In this result the propagators in the $B_0$ functions have the usual powers
 $\nu_i=1$ except when indicated $\nu_1=2$.
One can now even take the limit $m_1=0$ and then expand the result in $\delta$.
In section 4 more details on the products of $B_0$ functions and
their expansion will be given.

{}From eq. (\ref{T11234}) one can obtain special cases. Taking $m_1=0$
and one other mass vanishing double series are obtained.
In some cases one may obtain known hypergeometric functions, e.g. for
$m_1=m_2=0$,
$m_3=m_4=m$, which leads to
\begin{eqnarray}
&& T_{11234}(p^2;0,0,0,m^2,m^2) =
           2 \left(\frac{m^2}{4\pi\mu^2}\right)^{- 2\delta}
           \frac{\Gamma(\delta-1) \Gamma(1+2 \delta) \Gamma(3+\delta)}
                {\Gamma(5+2 \delta)}  \nonumber \\
&& \times {}_3 F_2 (1,\delta,2 \delta+1;2-\delta,\delta+5/2;\frac{p^2}{4 m^2})
   \nonumber\\
&&  + \left(\frac{m^2}{4\pi\mu^2}\right)^{- \delta}
     \left(\frac{-p^2}{4\pi\mu^2}\right)^{- \delta}
     \frac{\Gamma(1+\delta)\Gamma(\delta)\Gamma^2(1-\delta)}{\Gamma(1-2\delta)}
     \left(\frac{1}{6 m^2 (1-2\delta)} - \frac{1}{p^2 \delta} \right)\,.
\end{eqnarray}
\section{Analytic approaches and elliptic integrals}
In this section we inspect the imaginary parts of
the London transport diagram $T_{123}$ and of $T_{1234}$.
It turns out that they can be calculated in four dimensions
in terms of complete elliptic integrals. These are well known
functions and thus the results are of analytic interest.
Furthermore fast and precise algorithms for the calculation
of the elliptic integrals are available. Therefore the results provide
also an efficient way to calculate the imaginary parts numerically.

\subsection{Imaginary part of the London transport diagram}

As can be seen from (\ref{imag}) the imaginary part of
$T_{123}$ is convergent in four dimensions and reads with a
factorization of the K\"{a}ll\'{e}n functions
\begin{equation}
 Im(T_{123}(p^2))= \frac{1}{2 i} \Delta T_{123}(p^2) =
     -\frac{\pi}{p^2}
          \int\limits _{x_2} ^{x_3} { dt \over t }
          \sqrt { (t-x_1) (t-x_2) (x_3-t) (x_4-t) }\,,
          \label{ImT123short}
\end{equation}
with
\begin{eqnarray*}
 && x_1=(m_1-m_2)^2; \hskip3mm x_2=(m_1+m_2)^2; \hskip3mm x_3=(p-m_3)^2;
    \hskip3mm  x_4=(p+m_3)^2;  \\
 && x_1\le x_2 \le x_3 \le x_4; \hskip3mm p=\sqrt{|p^2|} \ge m_1+m_2+m_3.
\end{eqnarray*}
The integration limits are zeros of the square roots, and thus
(\ref{ImT123short}) leads to complete
elliptic integrals, defined by
\begin{eqnarray}
{\rm K}(x) &=& {\int\limits _0 ^1 {dt \over \sqrt{(1-t^2)(1- x ^2 t^2)}}}
 = {\pi \over 2}{_2 F _1}(-{1 \over 2},{1 \over 2};1;x^2)\,, \\
{\rm E}(x) &=& {\int\limits _0 ^1 {dt {(1- x^2 t^2)}
               \over \sqrt{(1-t^2)(1- x ^2 t^2)} }}
 = {\pi \over 2}{_2 F _1}({1 \over 2},{1 \over 2};1;x^2)\,, \\
\Pi(c,x) &=& {\int\limits _0 ^1 {dt  \over {(1-c t^2)
           \sqrt{(1-t^2)(1- x ^2 t^2)} }}}
   = {\pi \over 2} F_1 ({1 \over 2};1,{1 \over 2};1;c,x^2)\,,
\end{eqnarray}
with the Gauss hypergeometric function ${_2 F_1}$ and the Appell function
$F_1$ \cite{Exton,Appell,Srivastava}.
Reduction of (\ref{ImT123short}) to the Legendre normal form
of the elliptic integrals \cite{Erd,Abram}
by decomposition into partial fractions and partial integration
yields after some algebra
\begin{eqnarray}
 && Im \left( T_{123}(p^2;m_1^2,m_2^2,m_3^2) \right) = \nonumber \\
 && \;\;\; \;- \frac{\pi}{p^2} \bigg\{
 4m_1m_2 [(p+m_3)^2-m_3 p +m_1 m_2] \sqrt{q_{--} \over q_{++}}
   \; {\rm K}\left(\sqrt{\frac{q_{+-} q_{-+}}{q_{++} q_{--}}}\right)
  \nonumber \\
 && \;\;\;\;\;\;\;\;\;\;\;\;\; + {{m_1^2+m_2^2+m_3^2+p^2} \over 2} \sqrt{q_{++}
q_{--}}
 \; {\rm E}\left(\sqrt{\frac{q_{+-} q_{-+}}{q_{++} q_{--}}}\right)
 \nonumber \\
 && \;\;\;\;\;\;\;\;\;\;\;\;\; +{{8 m_1 m_2
[(m_1^2+m_2^2)(p^2+m_3^2)-2m_1^2m_2^2-2m_3^2p^2]}
         \over \sqrt{q_{++} q_{--}}}
   \; \Pi\left({q_{-+} \over q_{--}},
          \sqrt{\frac{q_{+-} q_{-+}}{q_{++} q_{--}}}\right)
 \nonumber \\
 && \;\;\;\;\;\;\;\;\;\;\;\;\; - {{8m_1m_2(p^2-m_3^2)^2} \over \sqrt{q_{++}
q_{--}}}
 \; \Pi\left({{{(m_1-m_2)^2}\over {(m_1+m_2)^2}} {q_{-+} \over q_{--}}},
            \sqrt{\frac{q_{+-} q_{-+}}{q_{++} q_{--}}}\right)
        \;\bigg\} \nonumber \\
   && \;\;\;\;\;\;\;\;\; \times \Theta \left( p^2-(m_1+m_2+m_3)^2 \right)\, ,
  \label{ImT123}
\end{eqnarray}
with variables $q_{\pm\pm}$ corresponding to the physical and
unphysical thresholds
\begin{equation}
q_{\pm \pm}:=(p \pm m_3)^2-(m_1 \pm m_2)^2.
\end{equation}

This result is valid in all parameter regions.
In special cases it leads to simpler formulae.
For equal masses one gets
\begin{eqnarray}
 && Im(T_{123}(p^2;m^2,m^2,m^2)) = - \frac{\pi}{p^2} \sqrt{(p-m)(p+3 m)}
       \nonumber \\
 && \hskip10mm \times
         \Big\{ -4 m^2 p \hskip2mm {\rm K}(\kappa)
              +\frac{(p-m)(p^2+3 m^2)}{2} \hskip2mm {\rm E}(\kappa) \Big\}
                  \,\Theta(p^2- 9 m^2), \\
 && \mbox{with} \;\; \kappa^2:=\frac{(p+m)^3(p-3 m)}{(p-m)^3(p+3 m)} \, ,
\end{eqnarray}
involving only complete elliptic integrals of the first and second kind, i.e.
${_2 F_1}$ Gauss hypergeometric functions.
If at least one mass is zero, $Im(T_{123})$ reduces to logarithms.
The most complicated case leads to
\begin{eqnarray}
 && Im(T_{123}(p^2;0,m_2^2,m_3^2)) =
  - \frac{\pi}{p^2} \bigg\{
 {{p^2+m_2^2+m_3^2} \over 2} \sqrt{q_+ q_-}
    \nonumber \\
  && \hskip12mm + (2 m_3^2 p^2 -m_2^2 (m_3^2+p^2) )
     \log ({{p^2+m_3^2-m_2^2-\sqrt{q_+ q_-}} \over {2 m_3 p}} )
   \nonumber \\
   && \hskip12mm + m_2^2 (p^2-m_3^2)
    \log({{2 m_2^2 m_3 p} \over {(p^2-m_3)^2-m_2^2(m_3^2+p^2)+(p^2-m_3^2)
            \sqrt{q_+ q_-}}})   \bigg\}
   \nonumber \\
  && \hskip8mm \times \Theta(p^2-(m_2+m_3)^2) \, ,
\end{eqnarray}
with $q_{\pm}:=(p \pm m_3)^2-m_2^2$.
\boldmath
\subsection{The imaginary part of $T_{1234}$}
\unboldmath

The two-particle cut contribution to the discontinuity
of $T_{1234}$ was given in (\ref{DeltaT12342}),
\begin{eqnarray}
 \Delta T^{(2)}_{1234}(p^2;m_1^2,m_2^2,m_3^2,m_4^2)
    &=& \Delta B_0(p^2;m_1^2,m_2^2) B_0(m_1^2+i \epsilon;m_3^2,m_4^2) \, .
 \label{DeltaT2}
\end{eqnarray}
As a product of a one-loop self-energy integral and a
one-loop self-energy discontinuity it is
composed of elementary functions and gets a real part for
\begin{equation}
  (m_3+m_4)^2 < m_1^2 \;\; \mbox{and} \;\; (m_1+m_2)^2 < p^2 \,.
  \label{m1diff}
\end{equation}
The three particle cut contribution can be calculated in a fashion
very similar to the case of the London transport diagram.
Only one more (complex conjugated) propagator
$1/(t-m_1^2-i \epsilon)$ has to be added in (\ref{ImT123short}).
The calculation yields
\begin{eqnarray}
 && \Delta T_{1234}^{(3)} (p^2;m_1^2,m_2^2,m_3^2,m_4^2) = \nonumber \\
 && \;\;\;\; \frac{2\pi i}{p^2}
 \bigg\{
4m_3m_4 \sqrt{q_{--} \over q_{++}}
   \; {\rm K}\left(\sqrt{\frac{q_{+-} q_{-+}}{q_{++} q_{--}}}\right)
 \; +  \sqrt{q_{++} q_{--}}
 \; {\rm E}\left(\sqrt{\frac{q_{+-} q_{-+}}{q_{++} q_{--}}}\right)
   \nonumber \\
 && \;\;\;\;\;\;\;\;\;\;\;\;\; +
   {{8 m_3 m_4 (p^2-m_1^2+m_2^2+m_3^2+m_4^2)}
         \over \sqrt{q_{++} q_{--}}}
   \; \Pi\left({q_{-+} \over q_{--}},
          \sqrt{\frac{q_{+-} q_{-+}}{q_{++} q_{--}}}\right)
 \nonumber \\
 && \;\;\;\;\;\;\;\;\;\;\;\;\; - {{8m_3m_4(p^2-m_2^2)^2} \over
{m_1^2\sqrt{q_{++} q_{--}}}}
 \; \Pi\left({{{(m_3-m_4)^2}\over {(m_3+m_4)^2}} {q_{-+} \over q_{--}}},
            \sqrt{\frac{q_{+-} q_{-+}}{q_{++} q_{--}}}\right)
   \nonumber \\
 && \;\;\;\;\;\;\;\;\;\;\;\;\; + {{8m_3m_4\lambda(p^2,m_1^2,m_2^2)}
                 \over {m_1^2\sqrt{q_{++} q_{--}}}}
 \; \Pi\left({{{m_1^2-(m_3-m_4)^2}
                 \over {m_1^2-(m_3+m_4)^2}} {q_{-+} \over q_{--}}}
                   -i\epsilon,
            \sqrt{\frac{q_{+-} q_{-+}}{q_{++} q_{--}}}\right)
        \bigg\}
    \nonumber \\
 && \;\;\;\; \times \Theta(p^2-(m_2+m_3+m_4)^2) \, ,
  \label{ImT1234}
\end{eqnarray}
with
\begin{equation}
q_{\pm \pm}:=(p \pm m_2)^2-(m_3 \pm m_4)^2.
\end{equation}

In the case (\ref{m1diff}) the characteristic c of the last $\Pi$-function
in (\ref{ImT1234}) is greater than $1$,
\begin{eqnarray}
  && c= {{{m_1^2-(m_3-m_4)^2}
          \over {m_1^2-(m_3+m_4)^2}} {q_{-+} \over q_{--}}} > 1 \, ,
\end{eqnarray}
which requires an analytic continuation of that function.
A comprehensive discussion of the analytic properties of
the elliptic integrals can be found in \cite{Tricomi}.
The $m_1^2+i \epsilon$ prescription in (\ref{ImT1234})
ensures that $\Delta T^{(3)}_{1234}$ gets the correct
real part, given through
\begin{eqnarray}
  Im \left( \Pi (c-i\epsilon,\kappa) \right) &=&
  {1 \over {2 i}} \left( \Pi(c-i\epsilon,\kappa)-\Pi(c+i\epsilon,\kappa)
             \right)
     \nonumber \\
  &=&   - \frac{ \pi}{2} \sqrt\frac{c}{(c-1)(c-\kappa^2)} \,.
\end{eqnarray}
This contribution cancels the real part of
the two-particle cut $\Delta T^{(2)}_{1234}$.
Consequently  $\Delta T_{1234}$ is always purely imaginary.

Numerical checks show the agreement of the results of (\ref{ImT123}) for
$Im(T_{123})$ and of
\begin{eqnarray}
  Im(T_{1234})=\frac{1}{2 i} \left( \Delta T^{(2)}_{1234}
  + \Delta T^{(3)}_{1234} \right)
\end{eqnarray}
with previously published tables \cite{Buza,Berends}.

\section{One-dimensional integral representations}

\subsection{A general approach to two-loop integrals containing
            a self-energy subloop}

An alternative method to the series expansion of the two-loop scalar
diagrams consists in the derivation of one-dimensional integral
representations. These are built up from one-loop self-energy functions
$B_0$ coming from the self-energy subloop and the remaining one-loop
integral. They can be derived by using a dispersion representation of
the $B_0$ function.

A two-loop diagram with only three-vertices
\vskip5mm
\beginpicture

\setcoordinatesystem units <.015mm,.015mm>
\setplotsymbol (.)

\circulararc 300 degrees from 866 500 center at 0 0
\circulararc 60 degrees from -866 -500 center at -1732 0

\plot 0 1000   0 1500 /
\plot 0 -1000   0 -1500 /
\plot 866 500  1299 750 /
\plot 866 -500  1299 -750 /

\put{$p_1$} [lt] at 0 -1500
\put{$p_2$} [lt] at 1299 -750
\put{$p_{N-1}$} [lb] at 1299 750
\put{$p_N$} [lb] at 0 1500
\put{$m_{N+2}$} [rt] at -500 -866
\put{$m_1$} [rb] at 500 -866
\put{$m_{N-1}$} [lb] at 500 866
\put{$m_{N+3}$} [rb] at -500 866
\put{$m_{N+1}$} [lB] at -700  0
\put{$m_{N}$} [rB] at -1050 0

\setdots
\circulararc 60 degrees from 866 -500 center at 0 0

\setplotsymbol ({\circle*{5}} [Bl])

\plot 866 500  866 501 /
\plot 866 -500  866 -501 /
\plot 0 1000  0 1001 /
\plot 0 -1000  0 -1001 /
\plot -866 500  -866 501 /
\plot -866 -500  -866 -501 /

\setplotsymbol (.)
\setsolid

\put{\Large{$=\frac{1}{m_{N+2}^2-m_{N+3}^2} $}\Huge{$\{$}} [lB] at 1200 0

\setcoordinatesystem units <.015mm,.015mm> point at -4800 0

\circulararc 300 degrees from 866 500 center at 0 0

\circulararc 90 degrees from -1000 0 center at -1000 1000

\plot 0 1000   0 1500 /
\plot 0 -1000   0 -1500 /
\plot 866 500  1299 750 /
\plot 866 -500  1299 -750 /


\put{\LARGE{$-$}} [lB] at 1100 0

\setdots
\circulararc 60 degrees from 866 -500 center at 0 0

\setplotsymbol ({\circle*{5}} [Bl])

\plot 866 500  866 501 /
\plot 866 -500  866 -501 /
\plot 0 1000  0 1001 /
\plot 0 -1000  0 -1001 /
\plot -1000 0  -1001 0 /

\setcoordinatesystem units <.015mm,.015mm> point at -7300 0
\setplotsymbol (.)
\setsolid

\circulararc 300 degrees from 866 500 center at 0 0

\circulararc 90 degrees from 0 -1000 center at -1000 -1000

\plot 0 1000   0 1500 /
\plot 0 -1000   0 -1500 /
\plot 866 500  1299 750 /
\plot 866 -500  1299 -750 /


\setdots
\circulararc 60 degrees from 866 -500 center at 0 0

\setplotsymbol ({\circle*{5}} [Bl])

\plot 866 500  866 501 /
\plot 866 -500  866 -501 /
\plot 0 1000  0 1001 /
\plot 0 -1000  0 -1001 /
\plot -1000 0  -1001 0 /

\put{\Huge{$\}$}} [lB] at 1200 0

\endpicture
\vskip5mm
where $k$ is the momentum flowing through the self-energy insertion,
can in a first step be reduced to simpler diagrams by a
decomposition into partial fractions
\begin{eqnarray}
 && \frac{1}{k^2-m_{N+2}^2} \frac{1}{k^2-m_{N+3}^2} =
    \frac{1}{m_{N+2}^2-m_{N+3}^2}
     \left( \frac{1}{k^2-m_{N+2}^2} - \frac{1}{k^2-m_{N+3}^2} \right)\,.
   \nonumber
\end{eqnarray}
This yields for the diagram
\begin{eqnarray}
  T_{1 \ldots N+3}(p_i;m_i^2) =
    \frac{1}{m_{N+2}^2-m_{N+3}^2}
    \bigg( &&\!\!\!\!\!\!\!\!
       T_{1 \ldots N+2}(p_i;m_1^2,\ldots,m_{N+1}^2,m_{N+2}^2)
    \nonumber \\
     &&\!\!\!\!\!\!\!\!
       - T_{1 \ldots N+2}(p_i;m_1^2,\ldots,m_{N+1}^2,m_{N+3}^2)
    \bigg)\,.
\end{eqnarray}
The difference has to be replaced by a derivative if $m_{N+2}^2=m_{N+3}^2$.

Insertion of the dispersion representation for the self-energy
subloop leads to
\begin{eqnarray}
  &&  T_{1 \ldots N+2}(p_i;m_1^2,\ldots,m_{N+1}^2,m_{N+2}^2) \nonumber \\
  && \;\;\; = \bigg< \, B_0(k^2;m_N^2,m_{N+1}^2)
               \, \frac{1}{(k+p_1)^2-m_1^2}
                \ldots \nonumber \\
  && \;\;\;\;\;\; \times \frac{1}{(k+p_1+\dots+p_{N-1})^2-m_{N-1}^2}
             \,  \frac{1}{k^2-m_{N+2}^2}  \bigg> \nonumber \\
  && \;\;\; = \frac{1}{2 \pi i} \int\limits_{s_0}^\infty ds \,
                 \Delta B_0(s;m_N^2,m_{N+1}^2)
               \bigg< \, \frac{-1}{k^2-s+i\epsilon} \nonumber \\
  && \;\;\;\;\;\; \times
                 \frac{1}{k^2-m_{N+2}^2}
                 \frac{1}{(k+p_1)^2-m_1^2} \ldots
                 \frac{1}{(k+p_1+\dots+p_{N-1})^2-m_{N-1}^2} \bigg>
\end{eqnarray}
with $s_0=(m_N+m_{N+1})^2$.
After a further decomposition into partial fractions,
\begin{eqnarray}
  &&  \frac{-1}{k^2-s} \;
      \frac{1}{k^2-m_{N+2}^2}
          = \frac{1}{s-m_{N+2}^2}
                \left(
                \frac{1}{k^2-m_{N+2}^2} - \frac{1}{k^2-s} \right)\,,
\end{eqnarray}
the $k$-integrations and one of the $s$-integrations
can be performed and yield
\begin{eqnarray}
  && T_{1 \ldots N+2}(p_i;m_1^2,\ldots,m_{N+1}^2,m_{N+2}^2) \nonumber \\
  && \;\;\; = B_0(m_{N+2}^2;m_N^2,m_{N+1}^2)
              T^{(1)}(p_i;m_1^2,\ldots,m_{N-1}^2,m_{N+2}^2) \nonumber \\
  && \;\;\;\;\;\; - \frac{1}{2 \pi i} \int_{s_0}^\infty ds
         \frac{\Delta B_0(s,m_N^2,m_{N+1}^2)}{s-m_{N+2}^2}
         T^{(1)}(p_i;m_1^2,\ldots,m_{N-1}^2,s)\,.
    \label{1disp}
\end{eqnarray}
$T^{(1)}$ denotes a one-loop N-point function in which $s$ enters
in the remaining one-dimensional integration as a mass variable.

A diagram with two four-vertices leads to a result which is
similar to the remaining integration in (\ref{1disp}),
\vskip3mm
\beginpicture

\setcoordinatesystem units <.01mm,.01mm>
\setplotsymbol (.)

\circulararc 300 degrees from 866 500 center at 0 0
\circulararc 60 degrees from -866 -500 center at -1732 0

\plot 866 500  1299 750 /
\plot 866 -500  1299 -750 /
\plot -866 500 -1299 750 /
\plot -866 -500 -1299 -750 /

\put{$p_1$} [lt] at -1250 -750
\put{$p_2$} [lt] at 1350 -750
\put{$p_{N-1}$} [lb] at 1350 750
\put{$p_{N}$} [lb] at -1250 750
\put{$m_1$} [lb] at -50 -900
\put{$m_{N-1}$} [lb] at -50 1000
\put{$m_{N}$} [lB] at -700  0
\put{$m_{N+1}$} [rB] at -1100 0

\put{ } [Bl] at -7000 0

\setdots
\circulararc 60 degrees from 866 -500 center at 0 0

\setplotsymbol ({\circle*{5}} [Bl])

\plot 866 500  866 501 /
\plot 866 -500  866 -501 /
\plot -866 500  -866 501 /
\plot -866 -500  -866 -501 /

\endpicture
\vskip3mm
\begin{eqnarray}
   T_{1 \ldots N+1}(p_i;m_i^2)
     &=& \frac{1}{2\pi i}
        \int \limits_{s_0}^\infty ds\, \Delta B_0(s;m_{N}^2,m_{N+1}^2)
        \nonumber \\
     && \;\;\; \times
        \bigg< \frac{-1}{k^2-s}
              \frac{1}{(k+p_1)^2-m_1^2} \ldots
        \frac{1}{(k+p_1+\ldots+p_{N-1})^2-m_{N-1}^2} \bigg>
         \nonumber \\
    &=&  - \frac{1}{2\pi i} \int \limits_{s_0}^\infty ds\,
        \Delta B_0(s;m_{N}^2,m_{N+1}^2)
        T^{(1)}(p_i;m_1^2,\ldots,m_{N-1}^2,s) \, .
 \label{2disp}
\end{eqnarray}
\subsection{Examples}

An application of (\ref{2disp}) to the London transport diagram
leads to
\begin{eqnarray}
  T_{123}(p^2;m_1^2,m_2^2,m_3^2) &=&
       - \frac{1}{2\pi i} \int \limits_{(m_2+m_3)^2}^\infty
   \!\!\!\! ds\,
        \Delta B_0(s;m_2^2,m_3^2) \,  B_0(p^2;s,m_1^2) \, ,
  \label{T123disp}
\end{eqnarray}
a result which would also follow from (\ref{disp}).
In that case a suitable subtraction \cite{Berends} is
\begin{eqnarray}
T_{123N}(p^{2};m_{1}^{2},m_{2}^{2},m_{3}^{2})
&=&
T_{123}(p^{2};m_{1}^{2},m_{2}^{2},m_{3}^{2})
- T_{123}(p^{2};m_{1}^{2},0,m_{3}^{2})
\label{defT123N}\\
&&
-T_{123}(p^{2};0,m_{2}^{2},m_{3}^{2})
+T_{123}(p^{2};0,0,m_{3}^{2})
\nonumber\,.
\end{eqnarray}

For  $T_{1234}$ one obtains from (\ref{1disp})
\begin{eqnarray}
  &&     T_{1234}(p^2;m_1^2,m_2^2,m_3^2,m_4^2) \nonumber \\
  &&\;\;\; = \frac{1}{2 \pi i} \int  \limits_{(m_3+m_4)^2}^\infty ds\,
        \frac{\Delta B_0(s;m_3^2,m_4^2)}{s-m_1^2+i\epsilon}
       \left( B_0(p^2;m_1^2,m_2^2) - B_0(p^2;s,m_2^2) \right)
  \label{T1234disp2}  \\
  &&\;\;\; =  B_0(m_1^2;m_3^2,m_4^2) B_0(p^2;m_1^2,m_2^2)  \nonumber \\
  &&\;\;\;\;\;\;  - \frac{1}{2\pi i} \int \limits_{(m_3+m_4)^2}^\infty ds\,
        \frac{\Delta B_0(s;m_3^2,m_4^2)}{s-m_1^2+i\epsilon}
         B_0(p^2;s,m_2^2)\,.
  \label{T1234disp}
\end{eqnarray}

The representations (\ref{T123disp}) and (\ref{T1234disp}) with the
subtractions (\ref{defT123N}) and (\ref{finite}) provide efficient ways to
calculate
$T_{123N}$ and $T_{1234N}$ in all parameter regions.
The results
agree numerically with those published in \cite{Buza}.

One may also consider vertex functions, for example

\vskip3mm

\beginpicture

\setcoordinatesystem units <.015mm,.015mm> point at 0 0
\setplotsymbol (.)

\circulararc 360 degrees from 866 500 center at 0 0

\circulararc 90 degrees from -1000 0 center at -1000 1000

\plot 0 1000   0 1500 /
\plot 0 -1000   0 -1500 /
\plot 1000 0  1500 0 /

\put{$p_1$} [lt] at 0 -1500
\put{$p_2$} [lB] at 1500 0
\put{$p_3=-p_1-p_2$} [lb] at 0 1500
\put{$m_2$} [lt] at 710 -710
\put{$m_1$} [rt] at -710 -710
\put{$m_3$} [lb] at 710 710
\put{$m_5$} [lt] at -300  150
\put{$m_4$} [rb] at -710 710

\put{ } [Bl] at -4000 0

\setplotsymbol ({\circle*{5}} [Bl])

\plot 0 1000  0 1001 /
\plot 0 -1000  0 -1001 /
\plot 1000 0  1001 0 /
\plot -1000 0  -1001 0 /

\endpicture

\begin{eqnarray}
  T(p_1^2,p_2^2,p_3^2;m_1^2,\ldots ,m_5^2) &=&
      B_0(m_1^2;m_4^2,m_5^2) \, C_0(p_1^2,p_2^2,p_3^2;m_1^2,m_2^2,m_3^2)
  \nonumber \\
 && - \frac{1}{2 \pi i} \int\limits_{s_0}^\infty ds\,
  \frac{\Delta B_0(s;m_4^2,m_5^2)}{s-m_1^2}
   \,  C_0(p_1^2,p_2^2,p_3^2;s,m_2^2,m_3^2)\,.
\end{eqnarray}
In that case the counterterm of the self-energy
subloop can be subtracted for a numerical evaluation.

In the cases of $T_{123N}$ and $T_{1234N}$ equivalent one-dimensional
integral representations can be obtained with a direct integration
in the momentum space or using the dispersion representation.


The case $T_{11234}$ for $m_1=0$ gets the representation
\begin{eqnarray}
&& T_{11234}(p^2;0,0,m_2^2,m_3^2,m_4^2) =
   B_0(p^2;\nu_1=2,\nu_2=1;0,m_2^2) B_0(0;m_3^2,m_4^2)
   \label{T11234int} \nonumber  \\
&& -\frac{1}{2 \pi i}
    \int\limits_{(m_{3}+m_{4})^2}^{\infty} \!\!\!\mbox{d} s\,\,
    \frac{\Delta B_0(s;m_3^2,m_4^2)}{s^2}
    \left[B_0(p^2;s,m_2^2)-B_0(p^2;0,m_2^2)\right]\,.
\end{eqnarray}
The integral is convergent and the result in an arbitrary number of
dimensions for the product $B_0 B_0$ which we denote by $Z$ is given by
\begin{eqnarray}
 Z &=&\frac{1}{m_2^2} \left(\frac{m_2^2}{4\pi\mu^2}\right)^{-\delta}
   \left(\frac{m_4^2}{4\pi\mu^2}\right)^{-\delta}
   \left(1-\frac{m_3^2}{m_4^2}\right)^{-1}
   \left[1- \left(\frac{m_3^2}{m_4^2}\right)^{1-\delta}\right]
   \nonumber \\
&& \times \frac{\Gamma^2(1+\delta)}{\delta^2 (1-\delta)^2}
   {}_2 F_1 (1+\delta,2;2-\delta;p^2/m_2^2).
\end{eqnarray}
One can perform the expansion of $Z$ in $\delta$ and obtains
\begin{eqnarray}
Z &=& \frac{(1-x)^{-1}}{m_2^2} \left(1- \frac{m_3^2}{m_4^2}\right)^{-1}
\left\{ \frac{1}{\delta^2} -\frac{1}{\delta}
\left(L_{m_{4}} + a + b -2 \right) + 3 + \zeta(2) + X \right. \nonumber \\
&& \left. +a \left(L_{m_{4}} + b -2\right) +
  \frac{1}{2}\left(L_{m_{4}} + b \right)^2
 - 2 \left(L_{m_{4}} + b\right) - \frac{m_3^2}{m_4^2} \left[L_{m_{4}}
  \rightarrow L_{m_{3}}\right] \right\} ,
\end{eqnarray}
where $x=p^2/m_2^2, a = 1 + \left(1/x-1\right) \ln (1-x)
,b=L_{m_{2}} + 2 \ln (1-x)$ and
\begin{equation}
X =  \left(\frac{1-x}{x}\right) \ln(1-x)
    -  \left(\frac{1-x}{x}\right) \ln^2(1-x)
    +  \left(\frac{1+x}{x}\right) \rm{Li_2}(x)\,.
\end{equation}

In the above representation masses can be set to zero, except the case
$m_3=m_4=0$. However, for this case an explicit result can be derived
\begin{eqnarray}
 T_{11234}(p^2;0,0,m^2,0,0) &=&
   \frac{1}{2 m^2}\left(\frac{m^2}{4\pi\mu^2}\right)^{-2 \delta}
   \frac{\Gamma(1+\delta) \Gamma^2(-\delta) \Gamma(1+2 \delta)}
        {(1-2 \delta) \Gamma(2-\delta)} \label{T11234(m2)} \nonumber  \\
&& \times {}_2 F_1 (1+2\delta,2+\delta;2-\delta;x)
\end{eqnarray}
with $x=p^2/m^2$.
Expanding the result around $D=4$ we get
\begin{eqnarray}
T_{11234}(p^2;0,0,m^2,0,0) &=&
\frac{1}{2 m^2 (1- x)}
\left\{ \frac{1}{\delta^2} -\frac{2}{\delta}\left(c+d-\frac{3}{2}\right)
+9 + 3 \zeta(2)  \right.  \nonumber\\
&& +2 d^2 -6 d + 4 c \left(d - \frac{3}{2}\right)+
4 \left(\frac{x-1}{x}\right) \ln^2(1-x)  \nonumber \\
&&\left.- 4 \left(\frac{x-1}{x}\right) \ln(1-x)
+ 2 \left(\frac{2 x +1}{x}\right) \rm{Li_2}(x) \right\}\, ,
\end{eqnarray}
where $ c = 1 + \left(1/x-1\right) \ln (1-x)$ and
 $d=L_m + 2 \ln (1-x)$.

For the other cases where some masses are zero the integral can also
be calculated analytically. A simple case is
\begin{eqnarray}
p^2 \, T_{11234} \! \left(p^2;0,0,0,m^2,0\right)&\!\!\!\!=&\!\!\!\!
- \frac{1}{\delta^2} + \frac{1}{\delta} \left\{ -1 + L_p + L_m \right\}
\! - \frac{3}{2} \! - \frac{1}{2} \, {\left( L_p + L_m \right)}^2
  \nonumber \\
&& + L_p + L_m
- \frac{1}{2} x \ln \left(-x\right)
 + \rm{Li_2}(x) \nonumber \\
&& + \frac{1}{2} \left( x- \frac{1}{x} \right) \ln \left( 1 - x \right)
\end{eqnarray}
where $x=p^2/m^2$, in agreement with \cite{Scharf}.

\section{Numerical comparisons and conclusions}
In this section numerical comparisons are made between the various
results of this paper and the literature.
Moreover some concluding remarks and an outlook are presented. Various
numerical
 evaluations for small $p^2$ are presented in Table 1.
In the first column are values
from the two-dimensional numerical integral representation \cite{Berends},
in the
second results from the series obtained from the hypergeometric functions,
(appendix), whereas column three uses three terms of the Taylor
expansion of \cite{Tausk} and finally the last column is calculated
from the one-dimensional integral of sect. 4.
The comparison gives confidence in
the various calculational methods.

In conclusion the analytic results in terms of generalized hypergeometric
functions offer the possibility to use well-known mathematical
techniques for analytic continuation, partial differential equations,
contour integral representations, etc.
 Moreover the result is derived for arbitrary masses from which one can get
many special cases. In other words the several formulas for the cases
with vanishing masses \cite{Scharf} are unified in one result in $D$
dimensions.
 When one is only interested in the imaginary parts an alternative analytic
result in four dimensions is obtained in terms of complete elliptic integrals.
Since their properties are well-known they are easily accessible
for numerical evaluations.
 Finally we have derived a one-dimensional integral representation for
all two-loop diagrams containing a self-energy insertion.
For the two-loop self-energy diagrams treated in this paper the integrand is
composed of elementary functions only and the representation
is valid for all values of $p^2$.
The main application of these integrals in this paper is for
numerical evaluations,
 giving a good alternative to the existing two-dimensional integrals. In order
 to complete the treatment of massive two-loop self-energy diagrams one has to
make the similar study of the master diagram.
As to the analytic approaches the prospects are not so good since
it is not of the self-energy insertion type.
In practice this means for instance that the massless case with arbitrary
powers
 of the propagators is not available in the literature.
This prevents an application of the Mellin-Barnes representation. On the other
hand symmetry properties, special cases and imaginary
parts have been studied in \cite{Broad}.
The two-dimensional integral representation was derived in particular for this
diagram \cite{Kreimer}. Nevertheless further studies will be needed.
Once adequate techniques for the self-energy diagrams are available one
could envisage practical applications for physics predictions.
For the electroweak theory the obvious application is to the
gauge boson self-energies which play a role in the
$M_W\,-\,M_Z$ mass relation and details of the Z line shape.
For QED the two-loop vacuum polarization has been known for a long time
\cite{Sabry}, but the electron two-loop self-energy was never fully calculated.

\begin{table}[htb]
\begin{center}
\begin{tabular}{|c|c|c|c|c|} \hline
$p^2$         & A             & B                & C             & D  \\
\hline
$2.66667$   & $-8.454752$   &  $-8.45044$   & $-8.445577$  & $-8.45038$ \\
$1.77778$   & $-8.286016$   &  $-8.28753$   & $-8.286116$  & $-8.28747$ \\
$1.18519$   & $-8.186514$   &  $-8.18486$   & $-8.184417$  & $-8.18481$ \\
$0.790124$  & $-8.116017$   &  $-8.11883$   & $-8.118665$  & $-8.11878$ \\
$0.526749$  & $-8.077571$   &  $-8.07583$   & $-8.075740$  & $-8.07577$ \\
$0.351166$  & $-8.053491$   &  $-8.04759$   & $-8.047528$  & $-8.04754$ \\
$0.234111$  & $-8.033367$   &  $-8.02896$   & $-8.028900$  & $-8.02890$ \\
$0.156074$  & $-8.013727$   &  $-8.01662$   & $-8.016561$  & $-8.01656$ \\
$0.104049$  & $-8.018667$   &  $-8.00843$   & $-8.008371$  & $-8.00837$ \\
$0.069366$  & $-7.999787$   &  $-8.00298$   & $-8.002926$  & $-8.00292$ \\
$0.046244$  & $-7.997535$   &  $-7.99936$   & $-7.999303$  & $-7.99930$ \\
$0.030829$  & $-7.985865$   &  $-7.99695$   & $-7.996891$  & $-7.99689$ \\
$0.020553$  & $-8.006871$   &  $-7.99534$   & $-7.995285$  & $-7.99528$ \\
$0.013702$  & $-7.900682$   &  $-7.99427$   & $-7.994214$  & $-7.99421$ \\
$0.009135$  & $-7.957410$   &  $-7.99356$   & $-7.993501$  & $-7.99350$ \\
$0.006090$  & $-7.793835$   &  $-7.99308$   & $-7.993026$  & $-7.99302$ \\
\hline
\end{tabular}
\end{center}
\caption[]{Real part of $T_{1234N}(p^2;m_1^2,m_2^2,m_3^2,m_4^2)$
for small values of
$p^2$. The masses are $m_1=1$, $m_2=3$, $m_3=5$ and $m_4=7$.
A: Result from VEGAS integration . B: Result of eqn.
(\ref{expan}).
C: Approximation using 3 terms of Taylor expansion.
D: One-dimensional integral representation (\ref{T1234disp}).}
\end{table}

\section*{Acknowledgement}
The authors of this paper greatfully acknowledge useful discussions
with J. B. Tausk, R. Scharf and G. Weiglein.

\section*{Appendix}

Here we give the result for the small $p^2$ expansion in $\delta$.
We introduce the dimensionless variables $x,y,u$ and $v$ defined by
\begin{equation}
 x=\frac{p^2}{m_2^2}\,,\, y=\frac{m_1^2}{m_2^2}\,,
\end{equation}
and
\begin{equation}
 u=\frac{m_1^2}{m_4^2}\,,\,v=\frac{m_3^2}{m_4^2}\,.
\end{equation}
We also need the roots $r_1$ and $r_2$ of the equation
\begin{equation}
 m_2^2 r + \frac{m_1^2}{r}=m_1^2 + m_2^2 -p^2
\end{equation}
and the roots $r_3$ and $r_4$ of the equation
\begin{equation}
 m_4^2 r + \frac{m_3^2}{r}=m_3^2 + m_4^2 - m_1^2\,.
\end{equation}

\begin{eqnarray}
\label{expan}
T_{1234N}(p^2;m_1^2,m_2^2,m_3^2,m_4^2) &=& -A1 + A2 + C1 + C2 + C3
                                       \nonumber \\
                                        && + C4 +C5 + C6 + C7
\end{eqnarray}
where
\begin{eqnarray}
 A1 &=&  8 - 2 \zeta(2)
   + \frac{1}{2} \left( 1 + \frac{y-1}{x} \right) \ln^2 (y)
  \nonumber \\
 &&
   + \left( -2 \left( 1 + \frac{y-1}{x} \right)
     - \frac{3}{4}  \frac{r_1-r_2}{x} \left( \ln(r_1) - \ln(r_2) \right)
     \right) \ln (y)
 \nonumber\\
 &&
   + \frac{1-x}{x} \ln (1-x)
   + \frac{1}{2} \left( \frac{1-y+x}{x} \right) \rm{Li_{2}}(x)
 \nonumber\\
 &&
   + \frac{1}{2} \frac{r_1-r_2}{x}
     \left\{ \vphantom{\left(\frac{-r_1(1-r_2)}{r_2-r_1}\right)}
     4 \ln(r_1) - 4 \ln(r_2)
\right.\nonumber\\   &&
     + \ln \! \left(\frac{1-r_1}{r_2-r_1}\right)
              \ln \! \left(\frac{r_1(1-r_2)}{r_1-r_2}\right)
  \! - \ln \! \left(\frac{1-r_2}{r_1-r_2}\right)
              \ln \! \left(\frac{r_2(1-r_1)}{r_2-r_1}\right)
 \nonumber\\
 &&
     + \rm{Li_{2}}\left(\frac{r_1(1-r_2)}{r_1-r_2}\right)
     - \rm{Li_{2}}\left(\frac{r_2(1-r_1)}{r_2-r_1}\right)
 \nonumber\\
 &&
     - \rm{Li_{2}}\left(\frac{1-r_2}{r_1-r_2}\right)
     + \rm{Li_{2}}\left(\frac{1-r_1}{r_2-r_1}\right)
     - \rm{Li_{2}}(1-r_1) + \rm{Li_{2}}(1-r_2)
 \nonumber\\
 &&
     - \rm{Li_{2}}\left(\frac{r_2(1-r_1)}{-r_1}\right)
     - \eta\left(1-x,\frac{1}{r_1}\right)
        \ln \left(\frac{r_2(1-r_1)}{-r_1}\right)
 \nonumber\\
 && \left.
     + \rm{Li_{2}}\left(\frac{r_1(1-r_2)}{-r_2}\right)
     + \eta\left(1-x,\frac{1}{r_2}\right)
        \ln \left(\frac{r_1(1-r_2)}{-r_2}\right)
\right\} .
\end{eqnarray}
\begin{eqnarray}
 &&A2 =
   \left\{
 -2 + \frac{1}{2} \frac{y-1}{x} \ln (y)
 - \frac{1}{2} \frac{r_1-r_2}{x}
   \left( \ln (r_1)
         -\ln (r_2) \right) \right\}
 \nonumber\\
 &&
  \times \left\{
 -2 + \frac{1}{2} \frac{v-1}{u} \ln (v)
 - \frac{1}{2} \frac{r_3-r_4}{u}
   \left( \ln (r_3)
         -\ln (r_4) \right) \right\}
 \nonumber\\
 &&
  +\frac{1}{2}\left\{
 -2 + \frac{1}{2} \frac{y-1}{x} \ln (y)
 - \frac{1}{2} \frac{r_1-r_2}{x}
   \left( \ln (r_1)
         -\ln (r_2) \right) \right\}
  \left\{\log\frac{m_1^2}{m_2^2}
        +\log\frac{m_3^2}{m_2^2}
        +\log\frac{m_4^2}{m_2^2}\right\}
  \nonumber\\
 &&
  +\frac{1}{2}\left\{
 -2 + \frac{1}{2} \frac{v-1}{u} \ln (v)
 - \frac{1}{2} \frac{r_3-r_4}{u}
   \left( \ln (r_3)
         -\ln (r_4) \right) \right\}
   \left\{\log\frac{m_1^2}{m_4^2}
        +\log\frac{m_2^2}{m_4^2}
        +\log\frac{m_3^2}{m_4^2}\right\}
  \nonumber\\
 &&
  +\frac{1}{8}\log^2\frac{m_3^2}{m_4^2}
  +\frac{1}{8}\log^2\frac{m_1^2}{m_2^2}
  +\frac{1}{4}\log\frac{m_1^2}{m_3^2}\log\frac{m_4^2}{m_2^2}
  +\frac{1}{4}\log\frac{m_1^2}{m_4^2}\log\frac{m_3^2}{m_2^2}
  +\log\frac{m_2^2}{m_4^2}
  \nonumber\\
 &&
  +\frac{1}{2}\left\{
 \zeta(2) + 8
 + \frac{1}{4} \ln^2 (y)
 \right. \nonumber\\
  &&
 - 2 \frac{y-1}{x} \ln (y)
 + \frac{r_1-r_2}{x}
 \left( \vphantom{\rm{Li_{2}}{ \frac{-r_1(1-r_2)}{r_2-r_1} } }
 2 \ln (r_1) - 2 \ln (r_2)
 \right. \nonumber\\
  &&
 + \ln \left( \frac{1-r_1}{r_2-r_1} \right)
   \ln \left( \frac{r_1(1-r_2)}{r_1-r_2} \right)
 - \ln \left( \frac{1-r_2}{r_1-r_2} \right)
   \ln \left( \frac{r_2(1-r_1)}{r_2-r_1} \right)
  \nonumber\\
  && \left. \left.
 + \rm{Li_{2}}\left( \frac{r_1(1-r_2)}{r_1-r_2}\right)
 - \rm{Li_{2}}\left( \frac{r_2(1-r_1)}{r_2-r_1}\right)
 - \rm{Li_{2}}\left( \frac{1-r_2}{r_1-r_2}\right)
 + \rm{Li_{2}}\left( \frac{1-r_1}{r_2-r_1}\right)
 \right)
 \right\}
 \nonumber\\
 &&
 +\frac{1}{2}\left\{
 \zeta(2) + 8
 + \frac{1}{4} \ln^2 (v)
 \right. \nonumber\\
  &&
 - 2 \frac{v-1}{u} \ln (v)
 + \frac{r_3-r_4}{u}
 \left( \vphantom{\rm{Li_{2}}{ \frac{-r_3(1-r_4)}{r_4-r_3} } }
 2 \ln (r_3) - 2 \ln (r_4)
 \right.\nonumber\\   &&
 + \ln \left( \frac{1-r_3}{r_4-r_3} \right)
   \ln \left( \frac{r_3(1-r_4)}{r_3-r_4} \right)
 - \ln \left( \frac{1-r_4}{r_3-r_4} \right)
   \ln \left( \frac{r_4(1-r_3)}{r_4-r_3} \right)
  \nonumber\\
  && \left. \left.
 + \rm{Li_{2}}\left( \frac{r_3(1-r_4)}{r_3-r_4}\right )
 - \rm{Li_{2}}\left( \frac{r_4(1-r_3)}{r_4-r_3}\right )
 - \rm{Li_{2}}\left( \frac{1-r_4}{r_3-r_4}\right )
 + \rm{Li_{2}}\left(  \frac{1-r_3}{r_4-r_3}\right )
 \right)
 \right\}.
\end{eqnarray}
\begin{eqnarray}
 \hskip-6mm
C1 &=& \sum_{\scriptstyle m,n=1\atop{\scriptstyle k,l=0}}^{\infty}
      \frac{(m+n)!(m+n-1)!(m+n+k+l-1)!(m+n+k+l)!}
         {m!(m+1)!n!(n-1)!l!(2(m+n+k)+l+1)!} \nonumber  \\
    && \times\left[-\psi(m+n) -2 \psi(m+n+k+l)
       -\psi(m+n+k+l+1)\right.\nonumber\\
    && \left.- \psi(m+2) + \psi(n) + 2 \psi(2(m+n+k)+2+l)\right]
       z_1^k z_2^n (1-z_3)^l z_4^m.
\end{eqnarray}
\begin{eqnarray}
 \!\!\!\!\!C2 &=&
    \sum_{m,n,k,l=0}^{\infty}
    \frac{(m+n)!(m+n+1)!(m+n+k+l+1)!(m+n+k+l)!}
         {m!(m+1)!n!(n+1)!l!(2(m+n+k)+l+3)!} \nonumber \\
    && \times\left[\log (m_4^2/m_2^2) + \psi(m+n+k+l+1)
       + \psi(m+2)\right.\nonumber\\
    && \left. + \psi(n+2) - \psi(m+n+2)\right]
       z_1^k z_2^{n+1} (1-z_3)^l z_4^m.
\end{eqnarray}
\begin{equation}
\hskip-24mm
    C3 =-
    \sum_{\scriptstyle k,l=0\atop{\scriptstyle m=1}}^{\infty}
    \frac{(m-1)!(m+k+l-1)!(m+k+l)!}
         {(m+1)!l!(2(m+k)+l+1)}
    z_1^k (1-z_3)^l z_4^m
\end{equation}
\begin{eqnarray}
 \!\!\!\!\!\!\!\!\!C4 &=&
    \sum_{\scriptstyle k,l=0\atop{\scriptstyle n=1}}^{\infty}
    \frac{(n+k+l-1)!(n+k+l)!}
         {l!(2(n+k)+l+1)!}
    z_1^k z_2^n (1-z_3)^l
    \nonumber \\
    && \left[-1 + 2 \psi(2(n+k)+l+2)
            - 2 \psi(n+k+l) -  \psi(n+k+l+1)\right].
\end{eqnarray}
\begin{eqnarray}
 \!\!\!\!\!\!C5 &=&
    \sum_{k,l=1}^{\infty}
    \frac{(k+l-1)!(k+l)!}
         {l!(2 k +l+1)!}
    z_1^k(1-z_3)^l
    \left[-1 - 2 \psi(k+l)\right. \nonumber \\
    &&\left. - \psi(k+l+1) + 2 \psi(2 k+l+2)\right].
\end{eqnarray}
\begin{equation}
 \!\!\!C6  =
    \sum_{k=1}^{\infty}
    \frac{k!(k-1)!}{(2 k+1)!}
    z_1^k\left[-1 - 2 \psi(k)
     - \psi(k+1) + 2 \psi(2 k +2)\right]
\end{equation}
\begin{equation}
 C7 =
    \sum_{l=1}^{\infty}
    \frac{(l-1)!}{(l+1)!}
    (1-z_3)^l \left[-1 - 2 \psi(l)
     - \psi(l+1) + 2 \psi(l+2)\right]
\end{equation}


\begin{thebibliography}{99}
\bibitem{Kreimer}D.~Kreimer,
{\em Phys.~Lett.~}B~273 (1991) 277.
\bibitem{Czarnecki}A.~Czarnecki, U.~Kilian and D.~Kreimer,
{\em Mainz preprint} MZ-TH/94-13.
\bibitem{Buza}F.~A.~Berends, M.~Buza, M.~B\"{o}hm and R.~Scharf,
{\em Z.~Phys.~}C~63 (1994) 227
\bibitem{Prudnikov} A.~P. Prudnikov, Yu.~A. Brychkov and
O.~I. Marichev,
{\em Integrals and Series}, Gordon and Breach Science
Publishers,
New York (1986).
\bibitem{Exton}H.~Exton,
{\em Multiple Hypergeometric Functions and Applications},
Ellis Horwood Limited,
 Chicester (1976).
\bibitem{Berends}F.~A.~Berends and J.~B.~Tausk,
{\em Nucl.~Phys.~}B~421 (1994) 456
\bibitem{Scharf}R.~Scharf and J.~B.~Tausk,
{\em Nucl.~Phys.~}B~412 (1994) 423
\bibitem{Smirnov}A.~I.~Davydychev, V.~A.~Smirnov and J.~B.~Tausk,
{\em Nucl.~Phys.~}B~410 (1993) 325.
\bibitem{Davyd}A.~I.~Davydychev and J.~B.~Tausk,
{\em Nucl.~Phys.~}B~397 (1993) 123.
\bibitem{Appell}P.~Appell, J.~Kamp\'{e} de F\'{e}riet,
{\em Fonctions Hyperg\'{e}om\'{e}triques}, Gauthier-Villars, Paris (1926).
\bibitem{Srivastava}H.~M.~Srivastava, P.~W.~Karlson,
{\em Multiple Gaussian Hypergeometric Series}, Halstead Press, John Wiley,
New York (1985).
\bibitem{Erd}A.~Erd\'{e}lyi, W.~Magnus, F.~Oberhettinger, F.~G.~Tricomi,
{\em Higher Transcendental Functions}, vol. II, McGraw-Hill,
New York (1953).
\bibitem{Abram}M.~Abramowitz, I.~A.~Stegun,
{\em Handbook of Mathematical Functions}, Dover Publications,
New York (1972).
\bibitem{Tricomi}F.~Tricomi, M.~Krafft,
{\em Elliptische Funktionen}, Akademische Verlagsgesellschaft,
Leipzig (1948).
\bibitem{Tausk}J.~B.~Tausk, {\em Top quark effects on multijet
production and electroweak parameters}, Leiden (1993).
\bibitem{Broad}D.~J.~Broadhurst, {\em Z.Phys.}, C 47 (1990) 115.
\bibitem{Sabry}G.~K\"{a}ll\'{e}n, A.~Sabry,
{\em Dan. Met. Fys. Medd.}, 29 (1955) no. 17
\end{thebibliography}
\end{document}